\documentclass[12pt]{article}
\newcommand{\bq}{\begin{equation}}
\newcommand{\eq}{\end{equation}}
\newcommand{\ba}{\begin{eqnarray}}
\newcommand{\ea}{\end{eqnarray}}

\newcommand{\nl }{ \nonumber  }

\newcommand{\p}{\partial}
\newcommand{\h}{\hspace{.5cm}}

\newcommand{\La}{\Lambda}

\begin{document}
\vspace*{1.cm}
\begin{center}
{\bf A NOTE ON TWO-SPIN MAGNON-LIKE ENERGY-CHARGE RELATIONS FROM
M-THEORY VIEWPOINT

\vspace*{1cm} P. Bozhilov} \footnote{e-mail:
bozhilov@inrne.bas.bg}

\ \\
\textit{Institute for Nuclear Research and Nuclear Energy,\\
Bulgarian Academy of Sciences, \\ 1784 Sofia, Bulgaria}

\end{center}
\vspace*{.5cm}
We show that for each M-theory background, having subspaces with
metrics of given type, there exist M2-brane configurations, which
in appropriate limit lead to two-spin magnon-like energy-charge
relations, established for strings on $AdS_5\times S^5$, its
$\beta$-deformation, and for membrane in $AdS_4\times S^7$.


\vspace*{.5cm} {\bf Keywords:} M-theory, M/field theory
correspondence, spin chains.

\section{Introduction}
One of the predictions of the AdS/CFT correspondence is that the
string theory on $AdS_5\times S^5$ should be dual to
$\mathcal{N}=4$ Super Yang Mills (SYM) theory in four dimensions
\cite{M97}, \cite{GKP98}, \cite{W98}. The spectrum of the string
states and of the operators in SYM should be the same. The recent
checks of this conjecture beyond the supergravity approximation
are connected to the idea to search for string solutions, which in
semiclassical limit (large conserved charges) are related to the
anomalous dimensions of certain gauge invariant operators in the
planar SYM \cite{BMN02}, \cite{GKP02}. On the field theory side it
was established that the corresponding dilatation operator is
connected to the Hamiltonian of integrable Heisenberg spin chain
\cite{MZ}.

In a recent paper \cite{HM}, Hofman and Maldasena explored a
specific semiclassical limit for strings on $R\times S^2$ subspace
of $AdS_5\times S^5$ and related it to the spin chain magnon
states. This limit leads to significant simplifications, and thus
allows for further improvement of our knowledge about the
string/gauge spectrum duality. More specifically, the "giant
magnon" solution obtained in \cite{HM} is a string with energy $E$
and spin $J$, which in the limit $E,J\to\infty$, $(E-J)$-finite,
obey the energy-charge relation \ba\nl
E-J=\frac{\sqrt{\lambda}}{\pi}\cos\theta_0,\ea where $\lambda$ is
the 't Hooft coupling, proportional to the square of the string
tension $T$, and the geometric angle $\theta_0$ is identified with
the magnon momentum $p$ on the spin chain side through the
equality \ba\nl \cos\theta_0=\mid\sin(p/2)\mid.\ea

In \cite{Dorey1}, N. Dorey proposed dispersion relation describing
magnon {\it bound states} \ba\label{mbs}
E-J=\sqrt{Q^2+\frac{\lambda}{\pi^2}\sin^2(p/2)},\ea where $Q$ is
the number of the constituent magnons, which should correspond on
the string theory side to the two-spin energy-charge relation
\ba\label{tsr1}
E-J_2=\sqrt{J_1^2+\frac{\lambda}{\pi^2}\sin^2(p/2)}.\ea The folded
string solution used in \cite{Dorey1}\footnote{Obtained in
\cite{FT03}, \cite{AFRT03}.} as confirmation of the above
proposal,  in the limit \ba\nl E, J_2\to\infty,\h E-J_2, J_1 -
\mbox{finite},\ea gives \ba\label{sr}
E-J_2=\sqrt{J_1^2+\frac{4\lambda}{\pi^2}}
=2\sqrt{\left(\frac{J_1}{2}\right)^2+\frac{\lambda}{\pi^2}}.\ea As
far as the folded string configuration is symmetric, this state
was interpreted as consisting of two excitations with momenta
$p=\pm \pi$, each carrying half of the total angular momentum
(spin) $J_1$. The conclusion drown was that then (\ref{sr}) agrees
with(\ref{mbs}). In a subsequent paper \cite{Dorey2}, N. Dorey et
al. was able to find string solution, which gives exactly the
relation (\ref{tsr1}) after the identification
$p=2\tan^{-1}(1/k)$, where $k$ is a free parameter. The same
result has been obtained in \cite{AFZ}-\cite{OS}, by identifying
different parameters in the string solutions with $p$, or by
purely group theoretic means \cite{CDO}. Evidently, the general
structure is \cite{MTT} \ba\label{gss} E-J_2=\sqrt{J_1^2+
k^2\lambda},\ea where $k$ is a constant depending on the
particular solution.

The above results have been obtained for strings moving on the
type IIB $AdS_5\times S^5$ background. However, it turns out that
relation of the type (\ref{gss}) also holds for strings on the
$\beta$-deformed $AdS_5\times S^5$ \cite{LM05}. The difference
with (\ref{tsr1}) is in the shift \ba\nl \frac{p}{2}\to
\frac{p}{2}-\pi\beta,\ea where $\beta$ is the deformation
parameter \cite{CGK06}, \cite{BobR06}.

The influence of the NS-NS field on the two-spin giant magnon has
been also examined \cite{WHH06}. The resulting changes in
(\ref{gss}) are: new constant $k^2$ and \ba\nl J_1^2\to
\mbox{const} J_1^2.\ea

For further investigations of the giant magnon properties see
\cite{MS}-\cite{BR0706} and references therein.

In this letter, we will show that there exist string
configurations, which satisfy magnon-like dispersion relations of
the type \ba\label{oss} E-AJ_2=\sqrt{BJ_1^2+ CT^2},\h T^2\sim
\lambda,\ea depending on three parameters $A$, $B$ and $C$.
Moreover, our main result is that the equality (\ref{oss}) also
holds for specific M2-brane configurations in M-theory. Such
solution has been already found for membrane moving on a subspace
of $AdS_4\times S^7$ \cite{BR06}.

\setcounter{equation}{0}
\section{Two-spin magnon-like relations from M-theory}
We will work with the following gauge fixed membrane action and
constraints \cite{NPB656}, which {\it coincide} with the gauge
fixed Polyakov type action and constraints after the
identification (see for instance \cite{27})
$2\lambda^0T_2=L=const$: \ba\label{omagf} &&S_{M}=\int d^{3}\xi
\mathcal{L}_{M} =\frac{1}{4\lambda^0}\int
d^{3}\xi\Bigl[G_{00}-\left(2\lambda^0T_2\right)^2\det
G_{ij}\Bigr],
\\ \label{00gf} &&G_{00}+\left(2\lambda^0T_2\right)^2\det G_{ij}=0,
\\ \label{0igf} &&G_{0i}=0.\ea In (\ref{omagf})-(\ref{0igf}), the
metric induced on the membrane worldvolume $G_{mn}$ is given by
\ba\label{im} &&G_{mn}= g_{MN}\p_m X^M\p_n X^N,\\ \nl
&&\p_m=\p/\p\xi^m,\h m = (0,i) = (0,1,2),\h M =
(0,1,\ldots,10),\ea where $g_{MN}$ is the target space metric. The
equations of motion for $X^M$, following from (\ref{omagf}), are
as follows $(\mathbf{G}\equiv\det{G_{ij}})$ \ba\label{eqm}
&&g_{LN}\left[\p_0^2 X^N - \left(2\lambda^0T_2\right)^2
\p_i\left(\mathbf{G}G^{ij}\p_j X^N\right)\right]\\ \nl
&&+\Gamma_{L,MN}\left[\p_0 X^M \p_0 X^N -
\left(2\lambda^0T_2\right)^2 \mathbf{G}G^{ij}\p_i X^M \p_j
X^N\right]=0,\ea where \ba\nl
\Gamma_{L,MN}=g_{LK}\Gamma^K_{MN}=\frac{1}{2}\left(\p_Mg_{NL}
+\p_Ng_{ML}-\p_Lg_{MN}\right)\ea are the components of the
symmetric connection corresponding to the metric $g_{MN}$.

If we split the target space coordinates as $x^M=(x^\mu, x^a)$,
where $x^\mu$ are those on which the background does not depend,
the conserved charges are given by the expression \cite{NPB656}
\ba\label{cc} Q_\mu =\frac{1}{2\lambda^0}\int d\xi^1 d\xi^2 g_{\mu
N}\p_0 X^N.\ea

Now, let us turn to our particular tasks. Consider backgrounds of
the type \ba\label{mtb}
ds^2=c^2\left[-dt^2+c_1^2d\theta^2+c_2^2\cos^2\theta d\varphi_1^2
+c_3^2\sin^2\theta
d\varphi_2^2+c_4^2f(\theta)d\varphi_3^2\right],\ea where $c$,
$c_1$, $c_2$, $c_3$, $c_4$ are arbitrary constants, and
$f(\theta)$ takes two values: $f(\theta)=1$ and
$f(\theta)=\sin^2\theta$. We embed the membrane into (\ref{mtb})
in the following way \ba\label{ra} &&X^0(\xi^m)\equiv t(\xi^m)=
\La_0^0\xi^0, \h X^1(\xi^m)=\theta(\xi^2),
\\ \nl &&X^2(\xi^m)\equiv \varphi_1(\xi^m)=
\La_0^2\xi^0,
\\ \nl &&X^3(\xi^m)\equiv \varphi_2(\xi^m)=
\La_0^3\xi^0,
\\ \nl &&X^4(\xi^m)=\varphi_3(\xi^m)=
\La_i^4\xi^i,
\\ \nl &&\mu=0,2,3,4,\h a=1,\h
\La_0^0,\ldots,\La_i^4=constants.\ea This ansatz corresponds to
M2-brane extended in the $\theta$- direction, moving with constant
energy $E$ along the $t$-coordinate, rotating in the planes
defined by the angles $\varphi_1$, $\varphi_2$, with constant
angular momenta $J_1$, $J_2$, and {\it wrapped} along $\varphi_3$.
The computations show that for this embedding, the constraints
$(\ref{0igf})$ and the equations of motion for the membrane
coordinates $X^\mu(\xi^m)$ are satisfied identically. Moreover, it
turns out that the remaining constraint (\ref{00gf}) is first
integral of the equation of motion for $X^a=X^1=\theta$
\cite{B0705}. That is why, it remains to solve the differential
equation (\ref{00gf}) only.

We begin with the case $f(\theta)=1$, when (\ref{00gf}) reduces to
\ba\label{ecp} &&K\theta'^2+V(\theta)=0,\\ \nl
&&K=-(2\lambda^0T_2c^2c_1c_4\La_1^4)^2,
\\ \nl &&V(\theta)=c^2\left\{(\Lambda_0^0)^2-(\Lambda_0^2c_2)^2
-\left[(\Lambda_0^3c_3)^2-(\Lambda_0^2c_2)^2\right]\sin^2\theta\right\}.\ea
From (\ref{ecp}) one obtains the turning point ($\theta'=0$) for
the effective one dimensional motion \ba\label{M^2}
M^2=\frac{(\Lambda_0^0)^2-(\Lambda_0^2c_2)^2}{(\Lambda_0^3c_3)^2-(\Lambda_0^2c_2)^2}.\ea
The solution of (\ref{ecp}) is \ba\label{s1} \xi^2(\theta)=
\frac{2\lambda^0T_2cc_1c_4\Lambda_1^4\sin\theta}{M\left[(\Lambda_0^3c_3)^2-(\Lambda_0^2c_2)^2\right]^{1/2}}
F_1(1/2,1/2,1/2;3/2;\sin^2\theta,\frac{\sin^2\theta}{M^2}),\ea
where $F_1(a,b_1,b_2;c;z_1,z_2)$ is one of the hypergeometric
functions of two variables \cite{PBM-III}. On this solution, the
conserved charges (\ref{cc}) take the form ($Q_0\equiv -E$,
$Q_2\equiv J_1$, $Q_3\equiv J_2$, $Q_4=0$) \ba\label{1E}
&&\frac{E}{\Lambda_0^0} =
\frac{2\pi^2T_2c^3c_1c_4\Lambda_1^4}{\left[(\Lambda_0^3c_3)^2-(\Lambda_0^2c_2)^2\right]^{1/2}}
\mbox{}_2F_1(1/2,1/2;1;M^2),
\\ \label{1J1} &&\frac{J_1}{\Lambda_0^2} =
\frac{2\pi^2T_2c^3c_1c_2^2c_4\Lambda_1^4}{\left[(\Lambda_0^3c_3)^2-(\Lambda_0^2c_2)^2\right]^{1/2}}
\mbox{}_2F_1(-1/2,1/2;1;M^2),
\\ \label{1J2} &&\frac{J_2}{\Lambda_0^3} =
\frac{2\pi^2T_2c^3c_1c_3^2c_4\Lambda_1^4}{\left[(\Lambda_0^3c_3)^2-(\Lambda_0^2c_2)^2\right]^{1/2}}
\left[\mbox{}_2F_1(1/2,1/2;1;M^2)-\mbox{}_2F_1(-1/2,1/2;1;M^2)
\right],\ea where $\mbox{}_2F_1(a,b;c;z)$ is the Gauss'
hypergeometric function.

Our next aim is to consider the limit, in which $M$ tends to its
maximum value: $M\to 1_-$. In this case, by using (\ref{M^2}) and
(\ref{1E})-(\ref{1J2}), one arrives at the energy-charge relation
\ba\label{2sgm} E-\frac{J_2}{c_3}=
\sqrt{\left(\frac{J_1}{c_2}\right)^2+\left(4\pi
T_2c^3c_1c_4\Lambda_1^4\right)^2} \ea for \ba\label{lim} E,
J_2/c_3\to\infty,\h E-J_2/c_3, J_1/c_2 - \mbox{finite}.\ea

Now, we are going to consider the case $f(\theta)=\sin^2\theta$
(see (\ref{mtb})), when (\ref{00gf}) takes the form
\ba\label{ecp1} &&\tilde{K}\theta'^2+V(\theta)=0,\\ \nl
&&\tilde{K}=-(2\lambda^0T_2c^2c_1c_4\La_1^4)^2\sin^2\theta
=K\sin^2\theta, \ea where $V(\theta)$ and correspondingly $M^2$
are the same as in (\ref{ecp}) and (\ref{M^2}). The solution of
(\ref{ecp1}) is given by the equality \ba\label{s2} \xi^2(\theta)=
\frac{\lambda^0T_2cc_1c_4\Lambda_1^4\sin^2\theta}
{M\left[(\Lambda_0^3c_3)^2-(\Lambda_0^2c_2)^2\right]^{1/2}}
F_1(1,1/2,1/2;2;\sin^2\theta,\frac{\sin^2\theta}{M^2}),\ea and is
obviously different from the previously obtained one. The
computations show that on (\ref{s2}) the conserved charges
(\ref{cc}) are as follows \ba\label{2E} &&\frac{E}{\Lambda_0^0} =
\frac{2\pi
T_2c^3c_1c_4\Lambda_1^4}{\left[(\Lambda_0^3c_3)^2-(\Lambda_0^2c_2)^2\right]^{1/2}}
\ln\left(\frac{1+M}{1-M}\right),
\\ \label{2J1} &&\frac{J_1}{\Lambda_0^2} =
\frac{2\pi
T_2c^3c_1c_2^2c_4\Lambda_1^4}{\left[(\Lambda_0^3c_3)^2-(\Lambda_0^2c_2)^2\right]^{1/2}}
\left[\frac{1-M^2}{2}\ln\left(\frac{1+M}{1-M}\right)+M\right],
\\ \label{2J2} &&\frac{J_2}{\Lambda_0^3} =
\frac{2\pi
T_2c^3c_1c_3^2c_4\Lambda_1^4}{\left[(\Lambda_0^3c_3)^2-(\Lambda_0^2c_2)^2\right]^{1/2}}
\left[\frac{1+M^2}{2}\ln\left(\frac{1+M}{1-M}\right)-M \right].\ea

Taking $M\to 1_-$, one sees that it corresponds again to the limit
(\ref{lim}), and the two-spin energy-charge relation is
\ba\label{2sgm'} E-\frac{J_2}{c_3}=
\sqrt{\left(\frac{J_1}{c_2}\right)^2+\left(2\pi
T_2c^3c_1c_4\Lambda_1^4\right)^2},\ea which differs from
(\ref{2sgm}) only by a factor of 4 in the second term on the right
hand side.

It is instructive to compare the above results with the string
case by using the same approach. To this end, for correspondence
with the membrane formulae, we will use the Polyakov action and
constraints in diagonal worldsheet gauge \ba\nl &&S_{S}=\int
d^{2}\xi \mathcal{L}_{S}= \int
d^{2}\xi\frac{1}{4\lambda^0}\Bigl[G_{00}-\left(2\lambda^0T\right)^2
G_{11}\Bigr],
\\ \nl &&G_{00}+\left(2\lambda^0T\right)^2 G_{11}=0,
\\ \nl &&G_{01}=0,\ea where
\ba\nl G_{mn}= g_{MN}\p_m X^M\p_n X^N,\h \p_m=\p/\p\xi^m, \h m =
(0,1),\h M = (0,1,\ldots,9).\ea The usually used conformal gauge
corresponds to $2\lambda^0T=1$.

An appropriate string theory background is \ba\label{stb}
ds^2=c^2\left[-dt^2+c_1^2d\theta^2+c_2^2\cos^2\theta d\varphi_1^2
+c_3^2\sin^2\theta d\varphi_2^2\right].\ea We consider string
embedding in (\ref{stb}) of the type \ba\nl &&X^0(\xi^m)\equiv
t(\xi^m)= \La_0^0\xi^0, \h X^1(\xi^m)=\theta(\xi^1),
\\ \label{se} &&X^2(\xi^m)\equiv \varphi_1(\xi^m)=
\La_0^2\xi^0,
\\ \nl &&X^3(\xi^m)\equiv \varphi_2(\xi^m)=
\La_0^3\xi^0, \h \La_0^0,\La_0^2,\La_0^3=constants.\ea This ansatz
corresponds to string extended in the $\theta$- direction, moving
with constant energy $E$, and rotating in the planes given by the
angles $\varphi_1$, $\varphi_2$, with constant angular momenta
$J_1$, $J_2$. Our calculations show that in the limit (\ref{lim}),
the string configuration (\ref{se}) is characterized by the
following magnon-like relation \ba\label{sml} E-\frac{J_2}{c_3}=
\sqrt{\left(\frac{J_1}{c_2}\right)^2+\left(4Tc^2c_1\right)^2}.\ea
Obviously, the two-spin energy-charge relations (\ref{2sgm}),
(\ref{2sgm'}) for membranes and (\ref{sml}) for strings are of the
same type.

\setcounter{equation}{0}
\section{Discussion}
We have shown here that for each M-theory background, having
subspaces with metrics of the type (\ref{mtb}), there exist
M2-brane configurations given by (\ref{ra}), which in the limit
(\ref{lim}) lead to the two-spin, magnon-like, energy-charge
relations (\ref{2sgm}) and (\ref{2sgm'}).

Examples for target space metrics of the type (\ref{mtb}) are
several subspaces of $R\times S^7$, contained in the $AdS_4\times
S^7$ solution of M-theory. As we already noticed in the
introduction, a membrane configuration has been found in
\cite{BR06}, corresponding to membrane moving on one of the
possible $AdS_4\times S^7$ subspaces, with the desired properties.
Namely, the background metric is given by \ba\nl
ds^2=(2l_pR)^2\left\{-dt^2+4\left[d\psi^2+\cos^2\psi
d\varphi_1^2+\sin^2\psi\left(\cos^2\theta_0
d\varphi_2^2+\sin^2\theta_0d\varphi_3^2\right)\right]\right\},\ea
where the angle $\theta$ is fixed to an arbitrary value
$\theta_0$, and the background 3-form field on $AdS_4$ vanishes.
The obtained two-spin, magnon-like, energy-charge relation is
\ba\nl E-\frac{J_2}{2\cos\theta_0}=
\sqrt{\left(\frac{J_1}{2}\right)^2+\left[2^6\pi T_2
(l_pR)^3\Lambda_1^4\sin\theta_0\right]^2},\ea and it corresponds
to $c=2l_pR$, $c_1=2$, $c_2=2$, $c_3=2\cos\theta_0$,
$c_4=2\sin\theta_0$ in (\ref{2sgm'}).

Moreover, it is not difficult to see that there exist 4 different
subspaces of $R\times S^7$ of the type (\ref{mtb}), when one of
the isometry coordinates $\varphi_1$, $\varphi_2$, $\varphi_3$ or
$\varphi_4$ equals zero, for which membrane embedding of the type
(\ref{ra}) ensures the existence of 12 solutions with
semiclassical behavior described by (\ref{2sgm}) or (\ref{2sgm'}),
corresponding to different values of the parameters $c$,
$c_1$,..., $c_4$.

Let us show that this is indeed the case. To this end, we
parameterize the metric on $R\times S^7$ subspace of $AdS_4\times
S^7$ as follows \ba\nl
ds^2&=&(2l_pR)^2\left\{-dt^2+4\left\{d\psi_1^2+\cos^2\psi_1
d\varphi_1^2\right.\right.\\ \nl
&+&\left.\left.\sin^2\psi_1\left[d\psi_2^2+\cos^2\psi_2
d\varphi_2^2+ \sin^2\psi_2\left(d\theta^2+\cos^2\theta
d\varphi_3^2+\sin^2\theta
d\varphi_4^2\right)\right]\right\}\right\}.\ea

If we fix $\varphi_4=0$, we will have two subcases for which the
metric will be of the type (\ref{mtb}): $(\psi_1,\theta)$ fixed to
$(\psi_1^0,\theta_0)$, \ba\nl
ds^2_1&=&(2l_pR)^2\left\{-dt^2+4\left[\cos^2\psi_1^0
d\varphi_1^2\right.\right.\\ \nl
&+&\left.\left.\sin^2\psi_1^0\left(d\psi_2^2+\cos^2\psi_2
d\varphi_2^2+ \sin^2\psi_2\cos^2\theta_0
d\varphi_3^2\right)\right]\right\},\ea and $(\psi_2,\theta)$ fixed
to $(\psi_2^0,\theta_0)$, \ba\nl
ds^2_2&=&(2l_pR)^2\left\{-dt^2+4\left[d\psi_1^2+\cos^2\psi_1
d\varphi_1^2\right.\right.\\ \nl
&+&\left.\left.\sin^2\psi_1\left(\cos^2\psi_2^0 d\varphi_2^2+
\sin^2\psi_2^0\cos^2\theta_0
d\varphi_3^2\right)\right]\right\}.\ea The appropriate membrane
embedding of the type (\ref{ra}) for the background given by
$ds_1^2$ is \ba\nl &&X^0(\xi^m)=t(\xi^m)= \La_0^0\xi^0, \\
\nl &&X^1(\xi^m)=\varphi_1(\xi^m)=\La_i^1\xi^i,
\\ \nl &&X^2(\xi^m)=\psi_2(\xi^2),
\\ \nl &&X^3(\xi^m)=\varphi_2(\xi^m)=
\La_0^3\xi^0,
\\ \nl &&X^4(\xi^m)=\varphi_3(\xi^m)=
\La_0^4\xi^0.\ea It corresponds to $J_{\varphi_1}=0$,
$(J_{\varphi_2},J_{\varphi_3})\ne 0$. In the limit $M\to 1_-$,
$J_{\varphi_2}$ is finite, whereas $J_{\varphi_3}\to\infty$. The
energy-charge relation $E(J_{\varphi_2},J_{\varphi_3})$ is
particular case of the one in (\ref{2sgm}), because $ds_1^2$
conform to $f=1$ in (\ref{mtb}). It reads \ba\nl
E-\frac{J_{\varphi_3}}{2\sin\psi_1^0\cos\theta_0}=
\sqrt{\left(\frac{J_{\varphi_2}}{2\sin\psi_1^0}\right)^2
+\left[2^7\pi
T_2(l_pR)^3\Lambda_1^1\sin\psi_1^0\cos\psi_1^0\right]^2}.\ea

For the background described by $ds^2_2$, there are two possible
embeddings of the type (\ref{ra}). They are
\ba\nl &&X^0(\xi^m)=t(\xi^m)= \La_0^0\xi^0, \\
\nl &&X^1(\xi^m)=\psi_1(\xi^2),
\\ \nl &&X^2(\xi^m)=\varphi_1(\xi^m)=\La_0^2\xi^0,
\\ \nl &&X^3(\xi^m)=\varphi_2(\xi^m)=
\La_0^3\xi^0,
\\ \nl &&X^4(\xi^m)=\varphi_3(\xi^m)=\La_i^4\xi^i
,\ea and
\ba\nl &&X^0(\xi^m)=t(\xi^m)= \La_0^0\xi^0, \\
\nl &&X^1(\xi^m)=\psi_1(\xi^2),
\\ \nl &&X^2(\xi^m)=\varphi_1(\xi^m)=\La_0^2\xi^0,
\\ \nl &&X^3(\xi^m)=\varphi_2(\xi^m)=\La_i^3\xi^i,
\\ \nl &&X^4(\xi^m)=\varphi_3(\xi^m)=\La_0^4\xi^0.\ea
For the first case, $(J_{\varphi_1},J_{\varphi_2})\ne 0$,
$J_{\varphi_3}=0$. In the limit $M\to 1_-$, $J_{\varphi_1}$ is
finite, while $J_{\varphi_2}\to\infty$. For the second case,
$(J_{\varphi_1},J_{\varphi_3})\ne 0$, whereas $J_{\varphi_2}=0$.
In the above mentioned limit, $J_{\varphi_1}$ is finite,
$J_{\varphi_3}\to\infty$. The energy-charge relations
$E(J_{\varphi_1},J_{\varphi_2})$ and
$E(J_{\varphi_1},J_{\varphi_3})$ are particular cases of the
relation (\ref{2sgm'}), because $ds_2^2$ correspond to
$f=\sin\theta$ in (\ref{mtb}). The explicit expressions for
$E(J_{\varphi_1},J_{\varphi_2})$ and
$E(J_{\varphi_1},J_{\varphi_3})$ are given by \ba\nl
E-\frac{J_{\varphi_2}}{2\cos\psi_2^0}=
\sqrt{\left(\frac{J_{\varphi_1}}{2}\right)^2 +\left[2^6\pi
T_2(l_pR)^3\Lambda_1^4\sin\psi_2^0\cos\theta_0\right]^2}\ea and
\ba\nl E-\frac{J_{\varphi_3}}{2\sin\psi_2^0\cos\theta_0}=
\sqrt{\left(\frac{J_{\varphi_1}}{2}\right)^2 +\left[2^6\pi
T_2(l_pR)^3\Lambda_1^3\cos\psi_2^0\right]^2}.\ea

Thus, we showed that for $\varphi_4=0$ there exist three membrane
configurations with the searched properties. By performing the
same analysis for the subspaces defined by $\varphi_1=0$,
$\varphi_2=0$ or $\varphi_3=0$, one can find another nine membrane
solutions with the same type of semiclassical behavior.

More examples for target space metrics of the type (\ref{mtb}),
for which there exist the membrane configurations (\ref{ra})
giving rise to two-spin magnon-like energy-charge relations, can
be found for instance in different subspaces of the $AdS_7\times
S^4$ solution of M-theory and not only there.

\vspace*{.5cm}

{\bf Acknowledgments} \vspace*{.2cm}

The author would like to thank R.C. Rashkov for an useful comment
on the subject considered here. This work is supported by NSFB
grants $\Phi-1412/04$ and $VU-F-201/06$.


\end{document}